%% file: IVvr23.tex
\documentclass[conference]{ieeeconf}
\usepackage[top=1.92cm, bottom=2.0cm, left=1.9455cm, right=1.9455cm]{geometry}
\IEEEoverridecommandlockouts
\usepackage{cite}
\usepackage{amsmath,amssymb,amsfonts}
\usepackage{algorithmic}
\usepackage{lipsum}
\usepackage{multirow}
\usepackage[table]{xcolor} 
\usepackage{graphicx}
\usepackage[caption=false]{subfig}
\usepackage{textcomp}
\usepackage{xcolor}
\usepackage{array}
\usepackage{booktabs}
\usepackage{makecell}
\usepackage{hhline}
\usepackage{soul}
\usepackage{capt-of}
\definecolor{rojo-anaranjado}{HTML}{FF5733}
\definecolor{azul-claro}{HTML}{AED6F1} 
\definecolor{azul-muy-claro}{HTML}{EAF6FF} 
\def\BibTeX{{\rm B\kern-.05em{\sc i\kern-.025em b}\kern-.08em
    T\kern-.1667em\lower.7ex\hbox{E}\kern-.125emX}}
\begin{document}


\title{\LARGE \bf Cross-cultural analysis of pedestrian group behaviour influence on crossing decisions in interactions with autonomous vehicles
} 






\author{S. Martín Serrano$^1$, O. Méndez Blanco$^1$, S. Worrall$^2$, M. A. Sotelo$^1$, D. Fern\'andez-Llorca$^{1,3}$
\thanks{$^{1}$Computer Engineering Department, Universidad de Alcal\'a, Spain.
}%
        \newline
\thanks{$^{2}$Australian Centre for Field Robotics, University of Sydney, Australia.     
}        
\newline
\thanks{$^{3}$European Commission, Joint Research Centre, Seville, Spain.     
}
}

\maketitle

\input{00_Abstract.tex}
\input{01_Introduction.tex}
\input{02_Method.tex}

\input{03_Results.tex}

\input{05_Conclusions.tex}

\section*{Acknowledgment}
We sincerely appreciate the participation of all individuals involved in the study.

\section*{Funding details}
This work was funded by Research Grants PID2020-114924RB-I00 and PDC2021-121324-I00 (Spanish Ministry of Science and
Innovation).  S. Martín Serrano acknowledges funding
from the University of Alcalá (FPI-UAH). D. Fernández Llorca acknowledges funding
from the HUMAINT project by the Directorate-General
Joint Research Centre of the European Commission. 

\section*{Disclaimer}
The views expressed in this article are purely those of the authors and may not, under any circumstances, be regarded as an official position of the European Commission.

\section*{Disclosure statement}
The authors declare that they have no competing interests to report.

\bibliographystyle{IEEEtran}
\bibliography{IVvr23}

\end{document}

%% file: 00_Abstract.tex
\begin{abstract}
Understanding cultural backgrounds is crucial for the seamless integration of autonomous driving into daily life as it ensures that systems are attuned to diverse societal norms and behaviours, enhancing acceptance and safety in varied cultural contexts. In this work, we investigate the impact of co-located pedestrians on crossing behaviour, considering cultural and situational factors. To accomplish this, a full-scale virtual reality (VR) environment was created in the CARLA simulator, enabling the identical experiment to be replicated in both Spain and Australia. Participants (N=30) attempted to cross the road at an urban crosswalk alongside other pedestrians exhibiting conservative to more daring behaviours, while an autonomous vehicle (AV) approached with different driving styles. For the analysis of interactions, we utilized questionnaires and direct measures of the moment when participants entered the lane. 

Our findings indicate that pedestrians tend to cross the same traffic gap together, even though reckless behaviour by the group reduces confidence and makes the situation perceived as more complex. Australian participants were willing to take fewer risks than Spanish participants, adopting more cautious behaviour when it was uncertain whether the AV would yield.

\end{abstract}


%% file: 01_Introduction.tex
\section{Introduction}

The integration of AVs into society depends on public opinion and acceptance, and there are numerous ongoing studies dedicated to exploring these aspects, along with examining cross-cultural variations and their impact \cite{Yun2021Cultural}. Various surveys in this regard report cultural differences in drivers' trust levels and their preferences for the design of AV explanation interfaces \cite{Hergeth2015,Na2023} or for vehicle interiors \cite{Vanessa2022}. The implementation of an accepted driving style is paramount to meet user expectations of safety and comfort. In addition to uncovering significant disparities between countries in how driving styles and decisions are interpreted \cite{Sun2023, Edelmann2021}, most of respondents also express a preference for more conservative AV driving behaviour compared to their own\cite{Tolbert2023}.

Another key factor in increasing the level of acceptance of AVs is to further our comprehension of their interactions with other road users, as the absence of a human driver prevents them from negotiating their actions in the conventional and socially compatible manner \cite{Wang2022}. From this arises the variety of Human-Machine Interfaces (HMIs) modalities \cite{bazilinskyy2019survey} and implicit communication strategies \cite{Tian2023} to convey their intentions and achieve their goals in the traffic scenarios. This information allows pedestrians gain confidence in AVs by being able to anticipate their movements and thus make more conscious and safer decisions when crossing streets.

Among the main variables on which pedestrians base their road-crossing decisions are the implicit communication with the vehicle over the explicit (i.e., legacy behaviour) \cite{Clamann2017} and the road-crossing situation, such as signage \cite{Leeknematics}, familiarity or neighbor's behaviour \cite{Anjara2023}. Even pedestrians who are not necessarily traveling together in a group can be influenced by seeing someone crossing the road and consequently modify their decision to cross. Social information plays an important role in decision-making that has also been observed to vary in the pedestrian context as a function of culture \cite{Pelé2017}.

Controlled virtual reality environments have demonstrated to offer optimal safety conditions and flexibility for the study of human-AV interaction \cite{MartinSerrano2024, Nascimento2019}. In this work, we conducted an experiment on this subject for which we employed a unique framework for the insertion of real agents into the CARLA simulator \cite{CarlaCHIRA2022, CarlaCHIRA23}. In it, an immersive VR interface and a full-body motion capture system allows a real pedestrian to interact with the simulation environment safely and adopting a natural locomotion style. We evaluate a form of implicit AV communication and the influence of co-located pedestrians on the user's decision making and subjective perception. For this purpose, we designed an interaction scenario in which an AV applied different braking manoeuvres as it approached a crosswalk, including occasions when it did not yield to the participant. A group of pedestrians stood at the beginning of the crosswalk and made crossing decisions together, which allowed us to answer our first research question: 

\begin{itemize}
    \item \textbf{RQ1}: \textit{To what extent do the braking manoeuvre and social information influence the crossing behaviour of a pedestrian in a crosswalk?}
\end{itemize}

Leveraging the high replicability of the simulated scenarios at minimal cost, we conducted the study in two different countries and analyzed the results of participants residing in Spain and Australia, thereby addressing our second research questions:

\begin{itemize}
    \item \textbf{RQ2}: \textit{Does cultural background influence the crossing behaviour of a pedestrian in a crosswalk?} 
\end{itemize}

Also, data captured in the VR environment and by the motion capture system can be used to train and test algorithms aimed at advancing autonomous driving technology.

\begin{figure}
\centerline{\includegraphics[width=0.89\columnwidth]{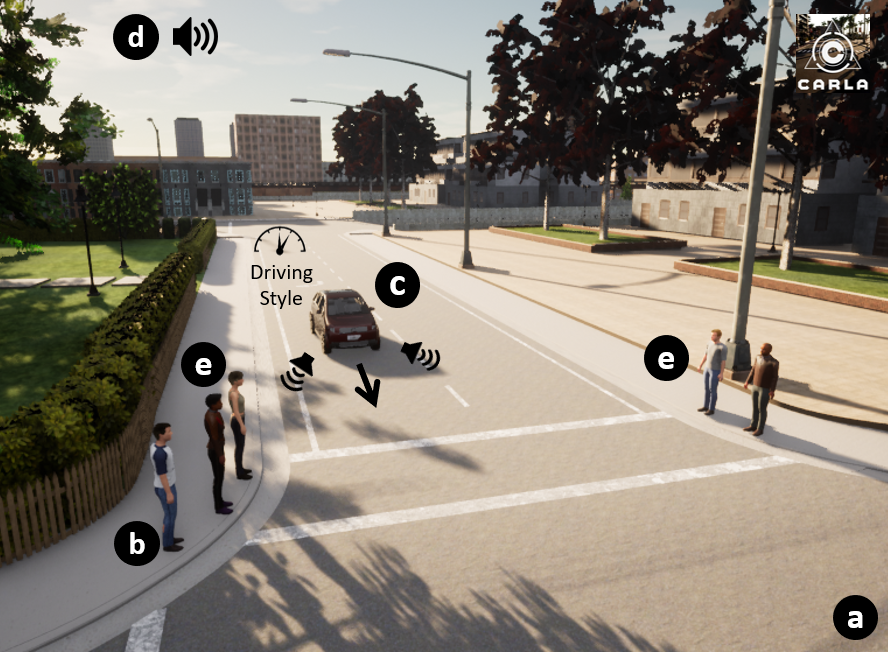}}
\caption{Experiment environment in CARLA. (a) 3D crosswalk scenario. (b) Participant matches the performer avatar. (c) Autonomous vehicle (engine sound, driving style). (d) Ambient sound, lighting and weather conditions. (e) Pedestrians attempting to cross from both sidewalks.}
\label{fig:schematic1}
\end{figure}

\section{Related work}

The influence of social information on pedestrian crossing behaviours has already been investigated in the literature suggesting that, if inadequate, it can lead to increased exposure to risk of injury \cite{Faria2010}. Collective behaviour has negative effects on road crossing safety. This work uses virtual reality and eye pattern scanning to reveal that the participants divert their attention to other pedestrians and assess danger less than if crossing alone \cite{Kwon2024}. Further, behavioural differences in pedestrian groups have been found to exist based on social interactions between group members (i.e., whether they are friends or strangers), gender and group size \cite{Barón2024, Ye2017}.

There is also evidence that context or country determine the effect of values on pedestrian behaviours \cite{Solmazer2020,McIlroy2020}. An apparent trend emerges wherein Asian participants exhibit a more cautious crossing behaviour compared to European participants, although it might not be as clear in route choices at crosswalks \cite{Sprenger2023}. Regarding the use of social information in different cultural contexts, this work shows that the Japanese were half as likely to be influenced by social information as the French when crossing at a red light, as they were more respectful of the rules \cite{Pelé2017}.

Research in multi-pedestrian scenarios has frequently employed virtual pedestrians, elevating interactions complexity, either in VR format \cite{Fratini2023, Stadler2019, Feng2023} or computer videos \cite{Hoggenmueller2022}. This enables us to incorporate controlled variables that program their behaviour, such as crossing the street early, disobeying AV communication, or standing still, and thus observe how human participants react to these situations. For example, this makes it possible to evaluate the impact of various eHMI designs in scenarios with multiple pedestrians, when a clearly defined recipient of the message is needed \cite{Dey2021projections}.

All previous works on this topic utilizing VR have been carried out in Unity, rather than in a specialized simulation environment designed specifically for autonomous driving, like the CARLA simulator we employ to advance the study of social information use from a cross-cultural perspective.

%% file: 02_Method.tex
\section{Method}

\subsection{Experiment Design}

The current study employed an immersive VR experiment to evaluate how social information, obtained from observing multiple co-located pedestrians, and the driving style of AVs influence pedestrian road crossing behaviour. A within-subject design approach was used whereby each participant was exposed to all experimental conditions to eliminate the effects of individual differences.

\subsubsection{Experiment Scenario Design}

Within one of the map models provided by the CARLA simulator \cite{CARLAalexey} we chose an urban crosswalk as the environment in which to conduct the VR experiment (see Fig. \ref{fig:schematic1}). Ambient audio was incorporated to make the VR experience more realistic, and the lighting and weather conditions were adjusted to be favorable. In this scenario, an AV drives down the street in a straight line and reaches the crosswalk when virtual pedestrians are waiting to cross from both sidewalks. The participant waits with their back to the road on one of the sidewalks and is instructed to turn around when the AV is 30 meters away and attempts to cross when deemed safe (no vegetation or other parked vehicles obstructing vision).

\begin{figure}
\centerline{\includegraphics[width=0.98\columnwidth]{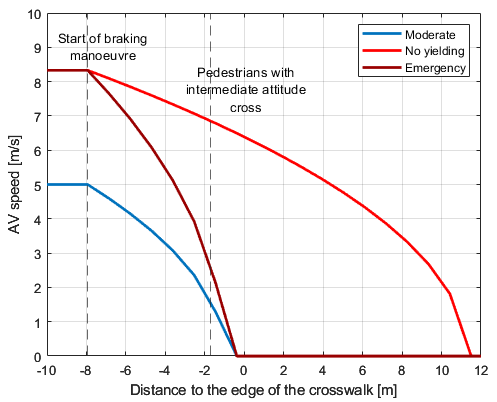}}
\caption{The three types of braking manoeuvre in the experiment: moderate, emergency and no yielding.}
\label{fig:varibles_at_crossing_gentle}
\end{figure}

As can be seen in Fig. \ref{fig:varibles_at_crossing_gentle}, three braking manoeuvres were designed to assess their effect on participant decision making. In the moderate manoeuvre, the AV travels comfortably at 18 km/h and applies a deceleration of -1.8 m/s\textsuperscript{2} until it stops before the crosswalk and yields to pedestrians. In the second manoeuvre, the AV circulates at 30 km/h and starts braking at the same point, so it has to apply a deceleration (emergency) of -5 m/s\textsuperscript{2} to come to a complete stop in order to also yield to pedestrians. Finally, in the third manoeuvre, the AV also travels at 30 km/h and begins to brake at the same distance as in the other cases but applying a deceleration of -1.8 m/s\textsuperscript{2}, thus it stops after passing the crosswalk.

\begin{figure}
\centering
\subfloat[]{\includegraphics[width=0.48\columnwidth]{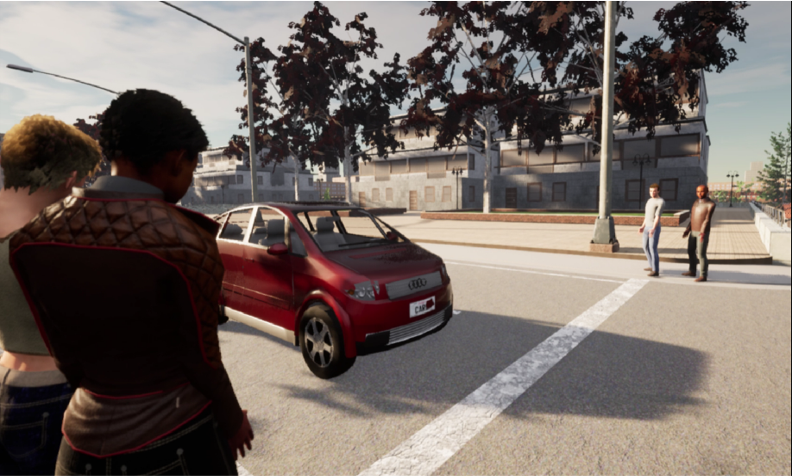}%
\label{fig:grail_green}}
\hfil
\subfloat[]{\includegraphics[width=0.48\columnwidth]{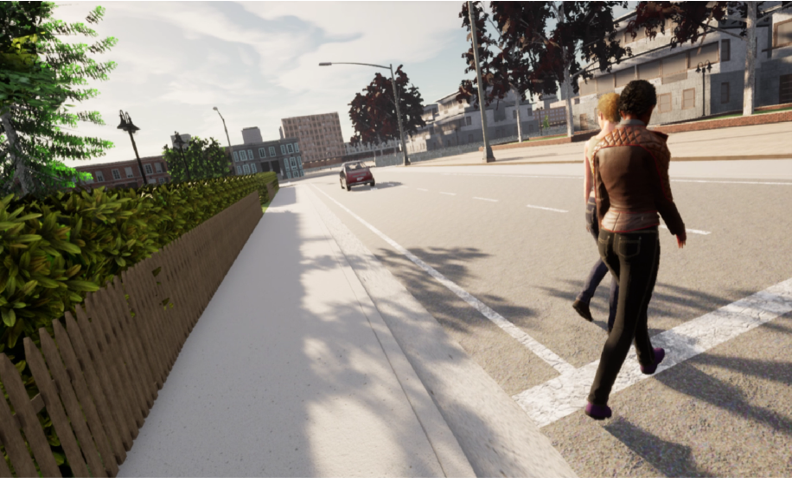}%
\label{fig:grail_red}}
\caption{(a) Pedestrians show a conservative attitude on both sidewalks and wait for the AV to come to a complete stop. (b) Pedestrians show a daring attitude and cross before the AV starts braking.  }
\label{fig:external_HMI}
\end{figure}

On the other hand, the behaviour of virtual pedestrians is modified to communicate different social information to the participant. As shown in Figures \ref{fig:grail_green} and \ref{fig:grail_red}, conservative pedestrians wait for the AV to fully decelerate  and pause for one second before crossing, whereas daring pedestrians cross when the AV is 25 meters away from the crosswalk, prior to the initiation of the braking manoeuvre. There's also a third intermediate behaviour where pedestrians enter the lane when the AV has covered 90\% of the braking distance (refer to Fig. \ref{fig:varibles_at_crossing_gentle}).

\subsubsection{Experiment Task Design}

The combination of different pedestrian attitudes and AV deceleration strategies result in the road-crossing tasks listed in Table \ref{tab:testsetup}. Moreover, Task 0 was added to familiarize participants with the environment, in which they were required to orient themselves towards the road in the absence of other pedestrians and simply observe the vehicle without crossing. The subsequent three tasks were completed in a randomized order tailored to each participant, introducing the types of braking manoeuvre. This acquainted participants with the possibility of the AV not yielding, even if they indicated their intention to cross (0-3 warm-up tasks).

The following seven tasks already counted for the drawing of conclusions and were also executed in a randomized order, when participants had to make the decision to cross in the face of different braking manoeuvres and virtual pedestrian attitudes. The last two tasks, also completed in a randomized order, combined pedestrians exposure to collision and failure to yield of the AV, recreating a hit-and-run situation (11-12 tasks).

\begin{table}[t]
\renewcommand{\arraystretch}{1.1}
\caption{Experimentation Tasks settings}
\begin{center}
\begin{tabular}{c|c|c|c|p{0.2cm}p{0.2cm}p{0.2cm}p{0.2cm}}
\textbf{Task}& \textbf{Braking} & \textbf{Pedestrians} & \textbf{Stop}\\
\textbf{Number}& \textbf{Manoeuvre} & \textbf{attitude} & \\
\hline
0   & -      & -    & No \\
\cline{1-4} 
1 & Moderate      & -     & Yes\\
2   & Emergency   & -     & Yes\\
3   & No yielding        & -     & Late\\
\hline
4   & Moderate    & Conservative & Yes\\
5   & Moderate    & Intermediate & Yes\\
6   & Moderate    & Daring & Yes\\
7   & Emergency    & Conservative & Yes\\
8   & Emergency    & Intermediate & Yes\\
9   & Emergency    & Daring & Yes\\
10   & No yielding    & Conservative & Late\\
\hline
11   & No yielding   & Intermediate & Late\\
12   & No yielding   & Daring & Late\\
\end{tabular}
\label{tab:testsetup}
\end{center}
\end{table}

\subsection{Virtual Reality Apparatus}

To facilitate the experiment within a VR environment, both in Spain and Australia we used the same framework for the insertion of real agents into CARLA \cite{CarlaCHIRA2022, CarlaCHIRA23}. Among the features added to the simulator were real-time avatar control, positional audio and body tracking. Two different models of head-mounted devices (HMD) were utilized: the Meta Quest 2 in Spain and the HTC Vive in Australia. The chosen full-body motion analysis system was XSens MVN \cite{ref_XSens2024} that also endows virtual pedestrians with natural motion dynamics. 

HMD was connected, wirelessly as a priority, to a Windows 10-11 desktop computer and an NVIDIA GeForce RTX 3060 graphics card. Through the VR interface, participants could engage as digited pedestrians within the CARLA traffic scenario along an 8 meter long x 3 meter wide space in which they adopted a real-walking locomotion style.

\subsection{Data Collection}

During the experiment, two types of data were collected to analyze the participant's interaction with the AV, including direct measurements (i.e., movement trajectory) and answers to questionnaires (i.e., subjective data). 

Within Unreal Engine project and MVN Analyze software \cite{CarlaCHIRA2022, CarlaCHIRA23}, all data generated during the VR experiment were logged, comprising: (1) timestamps, (2) vehicle positional and parameter data (i.e., x, y, z coordinates, rotation, brake, steer, throttle, gear), (3) participant and other pedestrians positional information and animations (i.e., x, y, z coordinates, rotation, .fbx), and (4) scene playback from the HDM view, the VR setup and the simulator, synchronized at 16 Hz. This allowed us to identify the moment when participants entered the vehicle lane, which we defined as the crossing event.

Between experimental tasks, participants had brief rest periods in which they did not leave the virtual environment when an accompanying researcher asked them the following questions about their subjective perception of the interaction they just had with the AV and their use of social information: 

\begin{enumerate}
\item[Q1:] How safe did you feel at the scene? 
\item[Q2:] How aggressive did you perceive the braking manoeuvre of the vehicle?
\item[Q3:] Did other pedestrians influence your decision to cross? 
\item[Q4:] Which pedestrians most influenced your crossing decision? None, those on your sidewalk, those on the opposite sidewalk or all of them indistinctly.  
\item[Q5:] How much mental and perceptual effort was required? Was the task easy or demanding, simple or complex?
\end{enumerate}

Answers to questions 1-3 were tabulated on a 7-step Likert scale \cite{joshi2015likert}. Question 5 (tabulated on a 20-step Likert scale) measures the cognitive and physical workload during user interaction with the VR environment to understand how it is affected by different experimental conditions.

\subsection{Participant’s Characteristics}

To collect data for additional cross-cultural analysis, we recruited participants from Alcalá de Henares (N = 15, Spain) and Sydney (N = 15, Australia) to engage in the experiment. We aimed for a similar age range and gender distribution in both groups. In Alcalá (ALC, M = 31.93, SD = 13.78), the gender distribution was 53$\%$ female and 47$\%$ male. In Sydney (SYD, M = 31.13, SD = 7.37), the gender distribution was 47$\%$ female and 53$\%$ male, with 60$\%$ of participants being of Asian origin. All of them had normal or corrected vision, normal mobility and reported no motion sickness.

In addition, we adhered to internal and institutional ethical review processes, which included notifying participants and obtaining their written consent, safeguarding data confidentiality and offering participants the option to withdraw from the experiments at any time and to use data anonymization.

%% file: 03_Results.tex
\section{Results}

This section presents the analysis of the impact of braking manoeuvre and social information on crossing decisions and questionnaire responses. To extract significant differences between the experimental tasks we employ the Student's t-test \cite{student} and the Wilcoxon signed-rank test \cite{woolson2007wilcoxon} respectively.

\subsection{Space Gap (L) results}

\begin{figure}
\centering
\renewcommand\thesubfigure{\roman{subfigure}}
\subfloat[]{\includegraphics[width=0.235\textwidth]{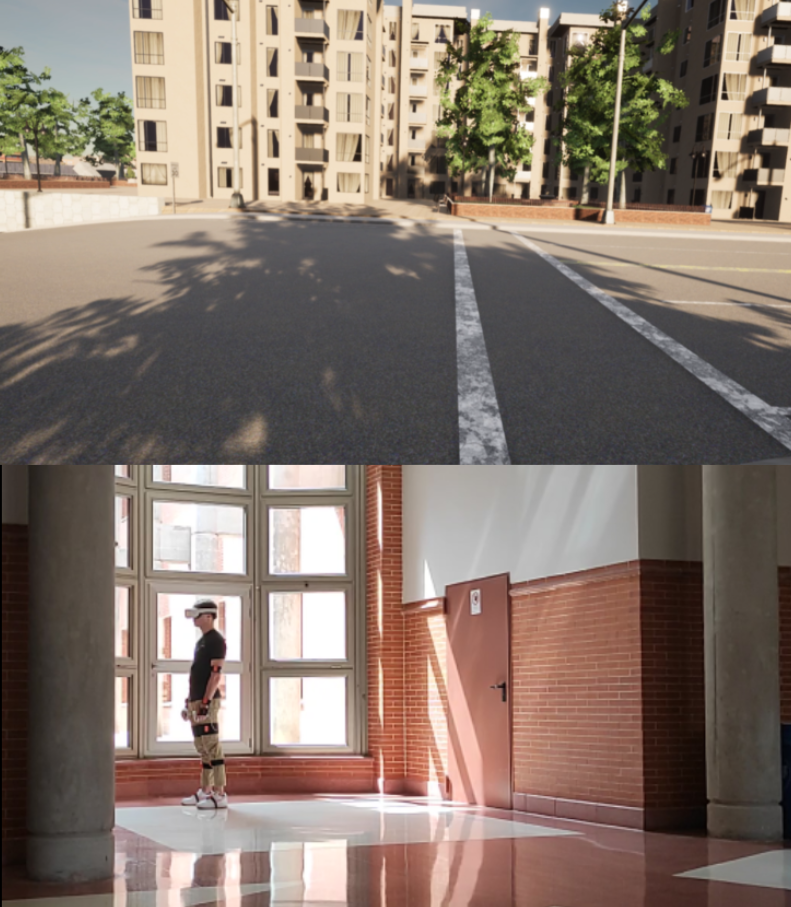}%
\label{fig_ped_1}}
\hfil
\subfloat[]{\includegraphics[width=0.235\textwidth]{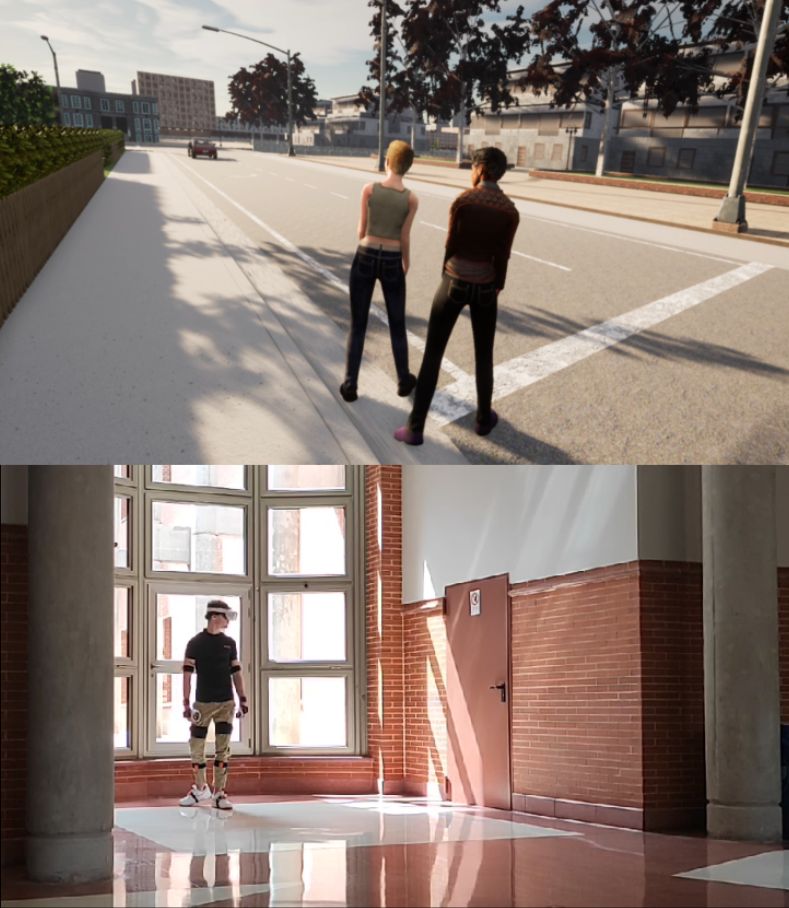}%
\label{fig_ped_2}}
\hfil
\subfloat[]{\includegraphics[width=0.235\textwidth]{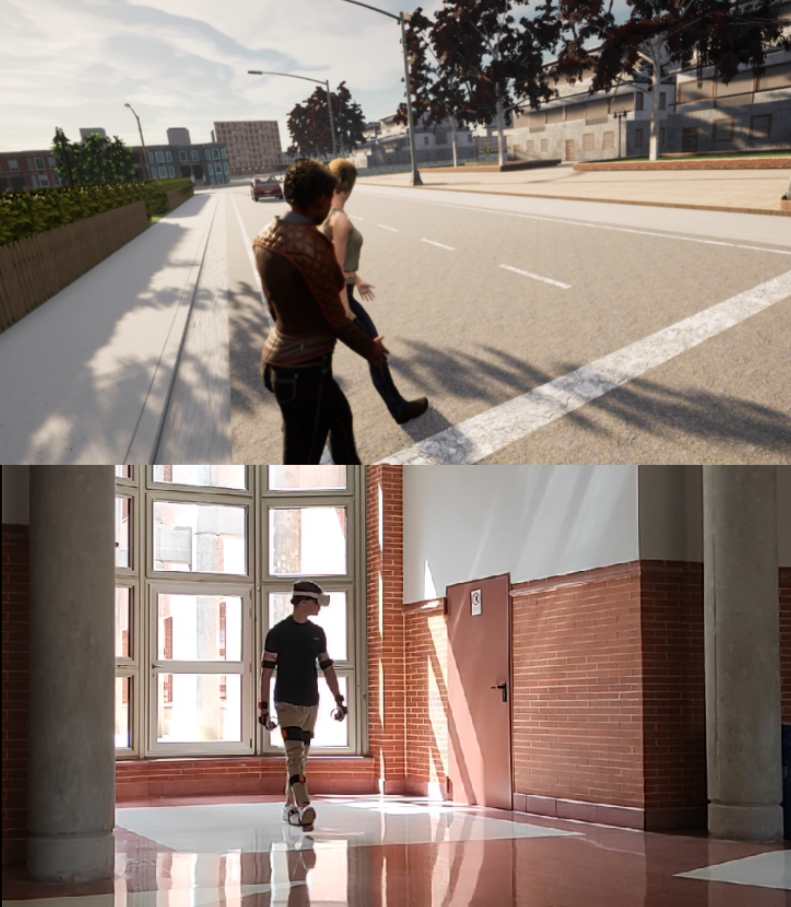}%
\label{fig_ped_3}}
\hfil
\subfloat[]{\includegraphics[width=0.235\textwidth]{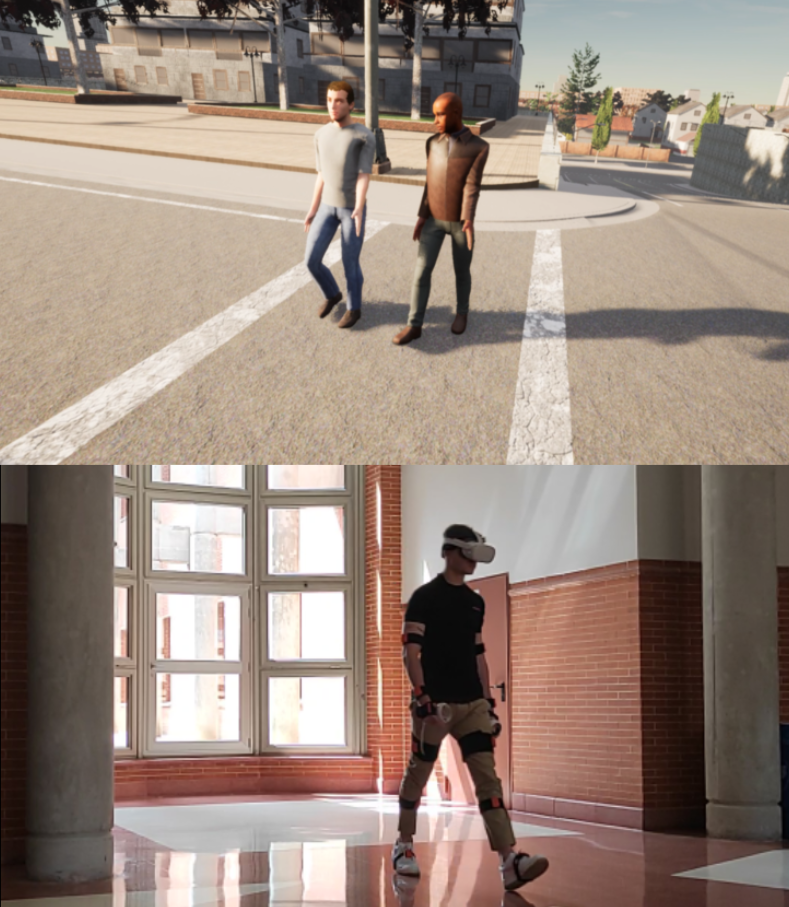}%
\label{fig_ped_4}}
\caption{Example of interaction between pedestrian and virtual AV with other pedestrians present: (i) Participant starts with their back to the crosswalk and hears a beep to turn around. (ii) Participant establishes visual contact with the AV and assesses the situation. (iii) Participant enters the lane together with the rest of the pedestrians, defining the crossing event. (iv) Participant crosses the road.}
\label{fig:example_interacion_exterior}
\end{figure}

\begin{figure}[t]
\centerline{\includegraphics[width=0.99\columnwidth]{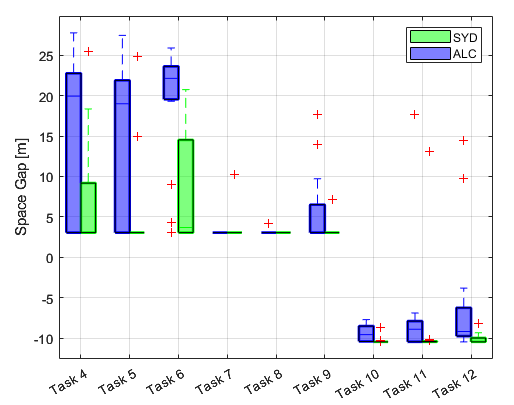}}
\caption{Box-plots of the pedestrian-AV distances at the crossing event. Performance of Spanish (ALC) and Australian (SYD) participants.}
\label{fig:boxplot_distance}
\end{figure}

The metric chosen to evaluate the crossing behaviour is the \emph{space gap} which is defined as the distance between the participant and the AV, measured from the AV to the centre of the crosswalk when the participant enters the lane and is exposed to a possible collision (see Fig. \ref{fig:example_interacion_exterior}). This informs us about their decision to proceed or delay crossing, as well as whether they crossed safely. Box-plots of the \emph{space gaps} in each task of the experiment are shown in Fig. \ref{fig:boxplot_distance}.

The Student’s t-test \cite{student} determines if there is a significant difference between the means of two related sample groups, considering both within-group variability and sample size. Table \ref{tab:student} expresses categorical statements, where a 1 in a cell denotes the rejection of the null hypothesis ($H_0:\mu_i\leq\mu_j$) and acceptance of the alternative hypothesis ($H_1:\mu_i>\mu_j$) with a confidence level of 90\%. This implies that the \emph{space gap} measured in task $i$ (row) is significantly larger than that in task $j$ (column).

Initially, it is notable Spanish participants tend to advance their crossing event when pedestrians show daring behaviour, even in instances where the AV does not yield  (L: t12 $>$ t10). In the case of Australians, this only occurs when applying the moderate manoeuvre (L: t6 $>$ t4, t5). A moderate manoeuvre consistently results in a larger \emph{space gap} than an emergency manoeuvre, while an emergency manoeuvre also leads to a larger \emph{space gap} compared to the no yielding situation.

The Table \ref{tab:L_gap} provides a direct comparison of \emph{space gaps} across both groups of different cultural backgrounds. The probability calculated by the Student’s t-test indicates that the alternative hypothesis is accepted in most variants of the experiment in favor of the Spanish participants. In essence, this implies that the Spanish exhibit more daring behaviour compared to Australians, as supported by the analysis of the \emph{space gap}.

\begin{table}[htbp]
\renewcommand{\arraystretch}{1.2}
\caption{Space Gap, Student t-test, $\alpha$=0.10}
\begin{center}
\begin{tabular}{ccc|p{0.2cm}p{0.2cm}p{0.2cm}p{0.2cm}p{0.2cm}p{0.2cm}p{0.2cm}p{0.2cm}p{0.2cm}}
\multicolumn{3}{c|}{\textbf{$H_1:\mu_i>\mu_j$}}& \multicolumn{9}{c}{\textbf{Task number $j$}} \\
 \multicolumn{3}{c|}{} & 4 & 5 & 6 & 7 & 8 & 9 & 10 & 11 & 12\\
\hline
\multirow{18}{*}{\rotatebox[origin=c]{90}{\textbf{Task number $i$}}} & \multirow{9}{*}{\rotatebox[origin=c]{90}{\textbf{L (ALC Score)}}}     

& 4   & \textbf{--}   & \textbf{0}  & \textbf{0}  & 1 & 1 & 1 & 1 & 1 & 1\\
&& 5   & \textbf{0}   & \textbf{--}  & \textbf{0}  & 1 & 1 & 1 & 1 & 1 & 1\\
&& 6   & \textbf{1}   & \textbf{1}  & \textbf{--}  & 1 & 1 & 1 & 1 & 1 & 1\\
&& 7   & 0   & 0  & 0  & \textbf{--}  & \textbf{0}  & \textbf{0}  & 1 & 1 & 1\\ 
&& 8   & 0   & 0  & 0  & \textbf{0}  & \textbf{--}  & \textbf{0}  & 1 & 1 & 1\\ 
&& 9   & 0   & 0  & 0  & \textbf{1}  & \textbf{1}  & \textbf{--}  & 1 & 1 & 1\\ 
&& 10  & 0   & 0  & 0  & 0 & 0 & 0 & \textbf{--} & \textbf{0} & \textbf{0}\\ 
&& 11  & 0   & 0  & 0  & 0 & 0 & 0 & \textbf{0} & \textbf{--} & \textbf{0}\\ 
&& 12  & 0   & 0  & 0  & 0 & 0 & 0 & \textbf{1} & \textbf{0} & \textbf{--}\\

\cline{3-12}
&\multirow{9}{*}{\rotatebox[origin=c]{90}{\textbf{L (SYD Score)}}} 

& 4   & \textbf{--}   & \textbf{0}  & \textbf{0}  & 1 & 1 & 1 & 1 & 1 & 1\\
&& 5   & \textbf{0}   & \textbf{--}  & \textbf{0}  & 1 & 1 & 1 & 1 & 1 & 1\\
&& 6   & \textbf{1}   & \textbf{1}  & \textbf{--}  & 1 & 1 & 1 & 1 & 1 & 1\\
&& 7   & 0   & 0  & 0  & \textbf{--}  & \textbf{0}  & \textbf{0}  & 1 & 1 & 1\\ 
&& 8   & 0   & 0  & 0  & \textbf{0}  & \textbf{--}  & \textbf{0}  & 1 & 1 & 1\\ 
&& 9   & 0   & 0  & 0  & \textbf{0}  & \textbf{0}  & \textbf{--}  & 1 & 1 & 1\\ 
&& 10  & 0   & 0  & 0  & 0 & 0 & 0 & \textbf{--} & \textbf{0} & \textbf{0}\\ 
&& 11  & 0   & 0  & 0  & 0 & 0 & 0 & \textbf{0} & \textbf{--} & \textbf{0}\\ 
&& 12  & 0   & 0  & 0  & 0 & 0 & 0 & \textbf{0} & \textbf{0} & \textbf{--}\\  
                
\end{tabular}
\label{tab:student}
\end{center}
\end{table}

\begin{table}[htbp]
\renewcommand{\arraystretch}{1.1}
\caption{Certainty of the discrepancy in Space Gap, Student t-test}
\begin{center}
\begin{tabular}{c|p{0.3cm}p{0.3cm}p{0.3cm}p{0.3cm}p{0.3cm}p{0.3cm}p{0.3cm}p{0.3cm}p{0.3cm}}
& \multicolumn{9}{c}{\textbf{Task number}}\\
$H_{1} (L)$: & \colorbox{gray!0}{4} & 5 & 6 & 7 & 8 & 9 & 10 & 11 & 12 \\
\hline
\rule{0pt}{10pt}$\mu_{alc}>\mu_{syd}$ & \textbf{1.00}  & \textbf{1.00} & \textbf{1.00}  & 0.16  & 0.84 & \textbf{0.96} & \textbf{1.00} & 0.74 & \textbf{0.98} \\

\rule{0pt}{10pt}$\mu_{syd}>\mu_{alc}$  & 0.00  & 0.00  & 0.00  & 0.84 & 0.16  & 0.04 & 0.00 & 0.26 & 0.02 \\

\end{tabular}
\label{tab:L_gap}
\end{center}
\end{table}

\subsection{Questionnaire results}

To evaluate the influence of the braking manoeuvre and social information on participants' questionnaire responses, we use the Wilcoxon signed-rank test \cite{woolson2007wilcoxon}, which serves as an alternative to the Student's t-test when dealing with data measured on ordinal or interval scales. Table \ref{tab:wilcoxon} uses the same null ($H_0:\mu_i\leq\mu_j$) and alternative hypotheses ($H_1:\mu_i>\mu_j$) to establish categorical statements about significant differences between experimental tasks at a 90\% confidence level. A 1 in a cell denotes rejection of the null hypothesis and acceptance of the alternative hypothesis, indicating that responses to a question in task $i$ (in the row) are significantly greater than those in task $j$ (in the column).

\begin{table}[htbp]
\renewcommand{\arraystretch}{1.1}
\caption{Wilcoxon signed rank test, Q1-3, $\alpha$=0.10}
\begin{center}
\begin{tabular}{ccc|p{0.2cm}p{0.2cm}p{0.2cm}p{0.2cm}p{0.2cm}p{0.2cm}p{0.2cm}p{0.2cm}p{0.2cm}}
\multicolumn{3}{c|}{\textbf{$H_1:\mu_i>\mu_j$}}& \multicolumn{9}{c}{\textbf{Task number $j$}} \\
 \multicolumn{3}{c|}{} & 4 & 5 & 6 & 7 & 8 & 9 & 10 & 11 & 12\\
\hline
\multirow{18}{*}{\rotatebox[origin=c]{90}{\textbf{Task number $i$}}} & \multirow{9}{*}{\rotatebox[origin=c]{90}{\textbf{Q1 (ALC Score)}}}     

& 4   & \textbf{--}   & \textbf{0}  & \textbf{0}  & 1 & 1 & 1 & 1 & 1 & 1\\
&& 5   & \textbf{0}   & \textbf{--}  & \textbf{0}  & 1 & 1 & 1 & 1 & 1 & 1\\
&& 6   & \textbf{0}   & \textbf{0}  & \textbf{--}  & 1 & 1 & 1 & 1 & 1 & 1\\
&& 7   & 0   & 0  & 0  & \textbf{--}  & \textbf{1}  & \textbf{1}  & 1 & 1 & 1\\ 
&& 8   & 0   & 0  & 0  & \textbf{0}  & \textbf{--}  & \textbf{0}  & 0 & 1 & 1\\ 
&& 9   & 0   & 0  & 0  & \textbf{0}  & \textbf{0}  & \textbf{--}  & 0 & 0 & 1\\ 
&& 10  & 0   & 0  & 0  & 0 & 0 & 0 & \textbf{--} & \textbf{0} & \textbf{1}\\ 
&& 11  & 0   & 0  & 0  & 0 & 0 & 0 & \textbf{0} & \textbf{--} & \textbf{0}\\ 
&& 12  & 0   & 0  & 0  & 0 & 0 & 0 & \textbf{0} & \textbf{0} & \textbf{--}\\ 

\cline{3-12}

& \multirow{9}{*}{\rotatebox[origin=c]{90}{\textbf{Q1 (SYD Score)}}}     

    & 4   & \textbf{--}   & \textbf{0}  & \textbf{1}  & 0 & 1 & 1 & 1 & 1 & 1\\
    && 5   & \textbf{0}   & \textbf{--}  & \textbf{1}  & 0 & 1 & 1 & 1 & 1 & 1\\
    && 6   & \textbf{0}   & \textbf{0}  & \textbf{--}  & 0 & 0 & 1 & 0 & 1 & 1\\
    && 7   & 0   & 0  & 0  & \textbf{--}  & \textbf{1}  & \textbf{1}  & 1 & 1 & 1\\ 
    && 8   & 0   & 0  & 0  & \textbf{0}  & \textbf{--}  & \textbf{0}  & 0 & 1 & 1\\ 
    && 9   & 0   & 0  & 0  & \textbf{0}  & \textbf{0}  & \textbf{--}  & 0 & 1 & 1\\ 
    && 10  & 0   & 0  & 0  & 0 & 0 & 0 & \textbf{--} & \textbf{1} & \textbf{1}\\ 
    && 11  & 0   & 0  & 0  & 0 & 0 & 0 & \textbf{0} & \textbf{--} & \textbf{0}\\ 
    && 12  & 0   & 0  & 0  & 0 & 0 & 0 & \textbf{0} & \textbf{1} & \textbf{--}\\

\cline{2-12}
\multirow{18}{*}{\rotatebox[origin=c]{90}{\textbf{Task number $i$}}} & \multirow{9}{*}{\rotatebox[origin=c]{90}{\textbf{Q2 (ALC Score)}}}  
                       & 4   & \textbf{--}   & \textbf{0}  & \textbf{0}  & 0 & 0 & 0 & 0 & 0 & 0\\
                      && 5   & \textbf{0}   & \textbf{--}  & \textbf{0}  & 0 & 0 & 0 & 0 & 0 & 0\\
                      && 6   & \textbf{0}   & \textbf{0}  & \textbf{--}  & 0 & 0 & 0 & 0 & 0 & 0\\
                      && 7   & 1   & 1  & 1  & \textbf{--}  & \textbf{0}  & \textbf{0}  & 0 & 0 & 0\\ 
                      && 8   & 1   & 1  & 1  & \textbf{0}  & \textbf{--}  & \textbf{0}  & 0 & 0 & 0\\ 
                      && 9   & 1   & 1  & 1  & \textbf{1}  & \textbf{0}  & \textbf{--}  & 0 & 0 & 0\\ 
                      && 10  & 1   & 1  & 1  & 1 & 1 & 1 & \textbf{--} & \textbf{0} & \textbf{0}\\ 
                      && 11  & 1   & 1  & 1  & 1 & 1 & 1 & \textbf{1} & \textbf{--} & \textbf{1}\\ 
                      && 12  & 1   & 1  & 1  & 1 & 1 & 1 & \textbf{0} & \textbf{0} & \textbf{--}\\ 
\cline{3-12}
&\multirow{9}{*}{\rotatebox[origin=c]{90}{\textbf{Q2 (SYD Score)}}} 
                       & 4   & \textbf{--}   & \textbf{0}  & \textbf{0}  & 0 & 0 & 0 & 0 & 0 & 0\\
                      && 5   & \textbf{1}   & \textbf{--}  & \textbf{0}  & 0 & 0 & 0 & 0 & 0 & 0\\
                      && 6   & \textbf{0}   & \textbf{0}  & \textbf{--}  & 0 & 0 & 0 & 0 & 0 & 0\\
                      && 7   & 1   & 1  & 1  & \textbf{--}  & \textbf{0}  & \textbf{0}  & 0 & 0 & 0\\ 
                      && 8   & 1   & 1  & 1  & \textbf{0}  & \textbf{--}  & \textbf{0}  & 0 & 0 & 0\\ 
                      && 9   & 1   & 1  & 1  & \textbf{1}  & \textbf{1}  & \textbf{--}  & 0 & 0 & 0\\ 
                      && 10  & 1   & 1  & 1  & 1 & 1 & 1 & \textbf{--} & \textbf{0} & \textbf{0}\\ 
                      && 11  & 1   & 1  & 1  & 1 & 1 & 1 & \textbf{1} & \textbf{--} & \textbf{0}\\ 
                      && 12  & 1   & 1  & 1  & 1 & 1 & 1 & \textbf{0} & \textbf{0} & \textbf{--}\\                       

\cline{2-12}
\multirow{18}{*}{\rotatebox[origin=c]{90}{\textbf{Task number $i$}}} & \multirow{9}{*}{\rotatebox[origin=c]{90}{\textbf{Q3 (ALC Score)}}}  
                       & 4   & \textbf{--}   & \textbf{1}  & \textbf{0}  & 0 & 0 & 0 & 0 & 0 & 0\\
                      && 5   & \textbf{0}   & \textbf{--}  & \textbf{0}  & 0 & 0 & 0 & 0 & 0 & 0\\
                      && 6   & \textbf{0}   & \textbf{1}  & \textbf{--}  & 0 & 0 & 0 & 0 & 0 & 0\\
                      && 7   & 0   & 1  & 0  & \textbf{--}  & \textbf{0}  & \textbf{0}  & 0 & 0 & 0\\ 
                      && 8   & 0   & 1  & 0  & \textbf{0}  & \textbf{--}  & \textbf{0}  & 0 & 0 & 0\\ 
                      && 9   & 1   & 1  & 0  & \textbf{1}  & \textbf{0}  & \textbf{--}  & 0 & 0 & 0\\ 
                      && 10  & 0   & 1  & 0  & 0 & 0 & 0 & \textbf{--} & \textbf{0} & \textbf{0}\\ 
                      && 11  & 0   & 1  & 0  & 0 & 0 & 0 & \textbf{0} & \textbf{--} & \textbf{0}\\ 
                      && 12  & 0   & 1  & 0  & 0 & 0 & 0 & \textbf{0} & \textbf{0} & \textbf{--}\\ 

\cline{3-12}
&\multirow{9}{*}{\rotatebox[origin=c]{90}{\textbf{Q3 (SYD Score)}}} 
                       & 4   & \textbf{--}   & \textbf{0}  & \textbf{0}  & 0 & 0 & 0 & 0 & 0 & 0\\
                      && 5   & \textbf{0}   & \textbf{--}  & \textbf{0}  & 0 & 0 & 0 & 0 & 0 & 1\\
                      && 6   & \textbf{0}   & \textbf{0}  & \textbf{--}  & 0 & 0 & 0 & 0 & 0 & 1\\
                      && 7   & 0   & 0  & 0  & \textbf{--}  & \textbf{0}  & \textbf{0}  & 0 & 0 & 0\\ 
                      && 8   & 0   & 0  & 0  & \textbf{0}  & \textbf{--}  & \textbf{0}  & 0 & 0 & 1\\ 
                      && 9   & 0   & 0  & 0  & \textbf{0}  & \textbf{0}  & \textbf{--}  & 0 & 0 & 0\\ 
                      && 10  & 0   & 0  & 0  & 0 & 0 & 0 & \textbf{--} & \textbf{0} & \textbf{0}\\ 
                      && 11  & 0   & 0  & 0  & 0 & 0 & 0 & \textbf{0} & \textbf{--} & \textbf{0}\\ 
                      && 12  & 0   & 0  & 0  & 0 & 0 & 0 & \textbf{0} & \textbf{0} & \textbf{--}\\

\end{tabular}
\label{tab:wilcoxon}
\end{center}
\end{table}

The first trend observed is that conservative attitude of the co-located pedestrians enhances participants' feeling of safety, with the exception of the Spanish when crossing in the face of a moderate manoeuvre who are not affected. Moderate deceleration increases the feeling of safety compared to emergency, as also does the emergency deceleration compared to not yielding, although not in all variations of pedestrians attitude. For the Spanish, there is no difference in their perception of safety whether pedestrians act daring during an emergency manoeuvre or do not cross when the AV does not yield (Q1: t9 vs t10). Australians also fail to discern a distinction when pedestrians are daring during a moderate manoeuvre compared to when they are conservative during an emergency manoeuvre (Q1: t6 vs t7).

Pedestrians crossing recklessly heighten the perception of aggressiveness associated with both the emergency manoeuvre and the failure-to-yield manoeuvre. For the Australians, additionally, an intermediate behaviour of nearby pedestrians makes the moderate deceleration perceived as more aggressive (Q2: t5 $>$ t4). The Spanish participants reported that pedestrians with an intermediate attitude during the moderate manoeuvre had minimal influence on them, while those with a daring attitude during the emergency manoeuvre exerted a somewhat greater impact, probably because they encouraged them to cross in a more precarious situation (Q3: t5, t9). Regarding Australian participants, they reported that pedestrians who crossed early when the AV did not yield had a slightly lower influence on their crossing decision than in some other tasks (Q3: t12). 

Table \ref{tab:wilcoxon_gap} presents direct comparisons between the questionnaire responses collected from the two groups of participants. The feeling of safety between Spanish and Australians in the face of the moderate braking is similar, although it diverges when pedestrians anticipate due to the fact that Spanish value it more positively. Australians felt safer with the emergency manoeuvre, and when other pedestrians were hit their confidence dropped dramatically. Spanish showed a greater appreciation for the distinction between types of manoeuvre, perceiving moderate braking as less aggressive, and emergency braking and no yielding as more aggressive. The Australians, moreover, reported feeling more influenced by nearby pedestrians, except when they demonstrated boldness with compromising manoeuvres (Q3: t9, t12).

\begin{table}[htbp]
\renewcommand{\arraystretch}{1.1}
\captionsetup{skip=-20pt}
\caption{Certainty of the discrepancy, Wilcoxon Signed Rank test}
\begin{center}
\begin{tabular}{c|p{0.3cm}p{0.3cm}p{0.3cm}p{0.3cm}p{0.3cm}p{0.3cm}p{0.3cm}p{0.3cm}p{0.3cm}}
& \multicolumn{9}{c}{\textbf{Task number}}\\
$H_{1} (Q1,2,3)$: & \colorbox{gray!0}{4} & 5 & 6 & 7 & 8 & 9 & 10 & 11 & 12 \\
\hline
\rule{0pt}{10pt}$\mu_{alc}>\mu_{syd}$  & 0.49  & 0.69  & \textbf{0.99}  & 0.07 & 0.07  & 0.20 & 0.13 & \textbf{1.00} & 0.40 \\

\rule{0pt}{10pt}$\mu_{syd}>\mu_{alc}$ & 0.51  & 0.31 & 0.01  & \textbf{0.93} & \textbf{0.93} & 0.80 & 0.87 & 0.00 & 0.60 \\

\hline
\rule{0pt}{10pt}$\mu_{alc}>\mu_{syd}$ & 0.32  & 0.07 & 0.05  & 0.86  & \textbf{0.97} & 0.58 & 0.82 & \textbf{0.93} & 0.88 \\

\rule{0pt}{10pt}$\mu_{syd}>\mu_{alc}$  & 0.68  & \textbf{0.93}  & \textbf{0.95}  & 0.14 & 0.03  & 0.42 & 0.18 & 0.07 & 0.12 \\

\hline
\rule{0pt}{10pt}$\mu_{alc}>\mu_{syd}$ & 0.23  & 0.00 & 0.04  & 0.07  & 0.11 & 0.85 & 0.22 & 0.46 & 0.73 \\

\rule{0pt}{10pt}$\mu_{syd}>\mu_{alc}$  & 0.77  & \textbf{1.00}  &  \textbf{0.96}  & \textbf{0.93} & 0.89  & 0.15 & 0.78 & 0.54 & 0.27 \\

\end{tabular}
\label{tab:wilcoxon_gap}
\end{center}
\end{table}

Table \ref{tab:q4_freq} summarizes in the form of a frequency table the choice responses to question 4, as well as the means of the responses to question 5. Most participants reported that the pedestrians who influenced them the most were those starting from their own sidewalk. The workload augmented as the manoeuvre threat escalated. Daring pedestrians reduced it during moderate braking, but increased it in the other cases.

\begin{table}[htbp]
\renewcommand{\arraystretch}{1.1}
\caption{Frequency Table for Q4 and Workload Q5}
\begin{center}
\begin{tabular}{c|p{0.3cm}p{0.3cm}p{0.3cm}p{0.3cm}p{0.3cm}p{0.3cm}p{0.3cm}p{0.3cm}p{0.3cm}}
& \multicolumn{9}{c}{\textbf{Task number}}\\
Q4-5 answers & 4 & 5 & 6 & 7 & 8 & 9 & 10 & 11 & 12 \\
\hline
\rule{0pt}{8pt}None - N         & 10 & 9   & 5   & 4   & 4  & 5   & 6   & 6   & 5 \\
Next to - NT     & 11 & 12  & 17  & 18  & 21 & 12  & 16  & 18  & 13 \\
In front - IF    & 1  & 1   & 1   & 2   & 1  & 4   & 0   & 2   & 1 \\
All - A          & 8  & 8   & 7   & 6   & 4  & 9   & 8   & 4   & 11 \\
Most chosen      & NT & NT  & NT  & NT  & NT & NT  & NT  & NT  & NT \\
\hline
\rule{0pt}{8pt}Work load      & 6.9 & 6.1  & 5.9  & 7.6  & 8.6  & 9.7  & 9.0  & 11.3  & 11.6 \\
SD      & 5.1 & 4.3  & 4.0  & 4.3  & 4.1  & 4.4  & 4.7  & 4.8  & 4.4 \\
\end{tabular}
\label{tab:q4_freq}
\end{center}
\end{table}

\subsection{RQ1 Discussion}

Co-located pedestrians cause the pedestrian under study to tend to cross alongside them even though, if the situation is perceived as threatening, their anticipation detracts from sense of safety, as well as amplifies the interaction perception as more dangerous and "complicated" (Q5). It is confirmed that inadequate social information could lead to exposure to injury (see t11, t12 in Fig. \ref{fig:boxplot_distance}). Participants clearly differentiated between the types of manoeuvres, though a conservative attitude of the nearby pedestrians proving to have a greater impact on confidence than the AV deceleration.

\subsection{RQ2 Discussion}

The Spanish participants were willing to take more risks than the Australian participants, even allowing themselves to be overly influenced by the pedestrian group in situations where the AV did not yield. The Australians adopted a more cautious attitude and valued emergency deceleration as safer than the Spanish  because of their internalized belief that they would only cross when there was a clear indication that the AV was going to stop. For the same reason, they reaffirmed that daring pedestrians did not advance their decision to cross because they did not follow them (Q3); instead, they waited to observe the vehicle's behaviour. Even so, like the Spanish, they also tended to cross together with nearby pedestrians when faced with a moderate type manoeuvre.

%% file: 05_Conclusions.tex
\section{Conclusions and Future Work}

We conducted a cross-cultural behaviour analysis of Spanish and Australian pedestrians within a VR environment. The findings reveal social information can modify road crossing decisions in the face of the same AV driving style. Further, they support previous assumptions about cultural differences in which pedestrians of Asian origin are more cautious than those of European origin. Among the limitations, this study took place in a a simple traffic scenario, with only one vehicle approaching and no other social activities in the background, except for the four pedestrians crossing from both sides of the street, as well as the influence of participants' past accident experiences was not considered. 

As future work, the behaviour of real pedestrians can be recreated inside the CARLA simulation to be detected by the virtual sensors equipped in the AV, including cameras, LiDAR, and radar. These synthetic sequences generated from the behaviour of real agents can serve as valuable resources for training and testing algorithms in the field of autonomous driving.